\documentclass[3p]{elsarticle}

\usepackage{graphics}
\usepackage{amssymb}
\def\Journal#1#2#3#4{{#1} {\bf #2}, (#4) #3}
\def\APJ{Astrophys. J.}

\def\EPJC{Eur. Phys. J. C}

\def\JETPUSSR{JETP (USSR)}
\def\JPG{J. Phys. G}

\def\NJP{New. J. Phys.}
\def\NPB{Nucl. Phys. B}

\def\NPBSUPPL{Nucl. Phys. B. Proc. Suppl.}
\def\PLB{{Phys. Lett.} B}

\def\PPNP{Prog. Part. Nucl. Phys.}

\def\PRL{Phys. Rev. Lett.}
\def\PRD{Phys. Rev. D}

\def\PTP{Prog. Theor. Phys.}
\def\RMP{Rev. Mod. Phys.}

\def\SCIENCE{Science}

\journal{Physics Letters B}

\begin{document}

\begin{frontmatter}

%% Title, authors and addresses

%% use the tnoteref command within \title for footnotes;
%% use the tnotetext command for the associated footnote;
%% use the fnref command within \author or \address for footnotes;
%% use the fntext command for the associated footnote;
%% use the corref command within \author for corresponding author footnotes;
%% use the cortext command for the associated footnote;
%% use the ead command for the email address,
%% and the form \ead[url] for the home page:
%%
%% \title{Title\tnoteref{label1}}
%% \tnotetext[label1]{}
%% \author{Name\corref{cor1}\fnref{label2}}
%% \ead{email address}
%% \ead[url]{home page}
%% \fntext[label2]{}
%% \cortext[cor1]{}
%% \address{Address\fnref{label3}}
%% \fntext[label3]{}

\title{Charged lepton contributions to bipair neutrino mixing}

%% use optional labels to link authors explicitly to addresses:
%% \author[label1,label2]{<author name>}
%% \address[label1]{<address>}
%% \address[label2]{<address>}

\author{Teruyuki Kitabayashi\corref{cor1}}
\ead{teruyuki@keyaki.cc.u-tokai.ac.jp}

\author{Masaki Yasu\`{e}}
\ead{yasue@keyaki.cc.u-tokai.ac.jp}

\cortext[cor1]{Corresponding author}

\address{Department of Physics, Tokai University, 4-1-1 Kitakaname, Hiratsuka, Kanagawa, 259-1292, Japan}

\begin{abstract}
The bipair neutrino mixing describes the observed solar and atmospheric mixings; however, it predicts vanishing reactor mixing angle, which is inconsistent with the observed data. We explore the ways of minimally modifying the bipair neutrino mixing by including charged lepton contributions. There are two categories of the bipair neutrino mixing which are referred to as case 1 and case 2. It turns out that, without arbitrary phases, a minimal modification is realized by just considering one $e$ - $\tau$ contribution from the charged lepton sector in the case 1. On the other hand, not only $e$ - $\tau$ contribution but also $\mu$ - $\tau$ contribution is required to realize a minimal modification in the case 2. 
\end{abstract}

\begin{keyword}
%% keywords here, in the form: keyword \sep keyword
bipair neutrino mixing \sep reactor mixing angle \sep charged lepton contributions
%% MSC codes here, in the form: \MSC code \sep code
%% or \MSC[2008] code \sep code (2000 is the default)

\end{keyword}

\end{frontmatter}

%%
%% Start line numbering here if you want
%%
% \linenumbers

%% main text
%%----------------------------------------------------------------------------------
\section{Introduction}
%%----------------------------------------------------------------------------------
The results from the neutrino oscillation experiments have provided us with robust evidence that neutrinos have tiny masses and their flavor states are mixed with each other \cite{atmospheric, solar, reactor, accelerator}. The recent global analysis \cite{global} shows that the best-fit values of the mixing angles are obtained as
\begin{eqnarray}
\sin^2 \theta_{12} &=& 0.312 \quad (0.280-0.347), \nonumber \\
\sin^2 \theta_{23} &=& 0.42  \quad (0.36-0.60),      \nonumber \\
\sin^2 \theta_{13} &=& 0.025 \quad (0.012 - 0.041),
\label{Eq:bestfit}
\end{eqnarray}
where $\theta_{12}, \theta_{23}$ and $\theta_{13}$ stand for the solar, atmospheric and reactor neutrino mixing angle, respectively. The values in the parentheses denote the $2\sigma$ range.

Various possibilities of realizing the observed mixing angles have been discussed \cite{reviewObMixings}. There is a theoretical prediction of these mixing angles based on the bipair neutrino mixing scheme \cite{bipair}, which yields
\begin{eqnarray}
\sin^2\theta_{12} = 0.293, \quad 
\sin^2\theta_{23} = 0.414, \quad
\sin^2\theta_{13} = 0,
\end{eqnarray}
for case 1 (BP1) and
\begin{eqnarray}
\sin^2\theta_{12} = 0.293, \quad
\sin^2\theta_{23} = 0.586, \quad
\sin^2\theta_{13} = 0,
\end{eqnarray}
for case 2 (BP2). There are other neutrino mixings that predict $\sin^2\theta_{13}=0$ such as the tribimaximal mixing (TB)  \cite{tribimaximal}, transposed tribimaximal mixing (TTB) \cite{transposedTribimaximal}, bimaximal mixing (BM) \cite{bimaximal}, two golden ratio schemes (GR1, GR2)  \cite{goldenRatio1, goldenRatio2} and hexagonal (HG) mixing \cite{hexagonal}. 

The predictability of the mixing scheme will be more transparent when it is expressed in terms of ratios:
\begin{eqnarray}
r_{ij} = \frac{\sin^2\theta_{ij}^{MS}}{\sin^2\theta_{ij}}, \quad (ij=12,23)
\end{eqnarray}
where $\theta_{ij}^{MS}$ in the numerator denotes the predicted mixing angle in the mixing scheme $MS$ =  $BP1$, $BP2$, $TB$, $TTB$, $BM$, $GR1$, $GR2$ and $HG$, while $\theta_{ij}$ in the denominator denotes the best-fit value of the observed mixing angle in Eq.(\ref{Eq:bestfit}). Numerically, we obtain 
\begin{eqnarray}
(r_{12}^{BP1}, r_{23}^{BP1}) = (0.94, 0.99), \quad
 (r_{12}^{BP2}, r_{23}^{BP2}) = (0.94, 1.4),
\end{eqnarray}
for the bipair neutrino mixing and 
\begin{eqnarray}
&& (r_{12}^{TB}, r_{23}^{TB}) = (1.06,1.19),  \quad
      (r_{12}^{TTB},r_{23}^{TTB}) = (1.60,1.59), \quad
      (r_{12}^{BM},r_{23}^{BM}) = (1.60,1.19), \nonumber \\
&& (r_{12}^{GR1},r_{23}^{GR1}) = (0.89,1.19),\quad
     (r_{12}^{GR2},r_{23}^{GR2}) = (1.11,1.19), \quad
     (r_{12}^{HG},r_{23}^{HG}) = (0.80,1.19),
\end{eqnarray}
for other mixings. The bipair neutrino mixing BP1 turns out to be a better candidate as far as the solar and atmospheric neutrino mixings are concerned.

Since the tribimaximal mixing scheme can arise from a certain symmetry argument, this scheme has been extensively studied \cite{reviewObMixings}. Although the bipair neutrino mixing scheme does not originate from any symmetry argument imposed either on the neutrino mass matrix or on the Lagrangian, it seems that the predicted values of the solar and atmospheric mixing angles in the case 1 of the bipair neutrino mixing as well as the tribimaximal mixing are remarkably well compatible with the observed data. Namely, the case 1 of the bipair  neutrino mixing becomes consistent with the recent global analysis of the neutrino mixings \cite{global} although another global analysis slightly disfavors the case 1 \cite{otherGlobal}.

The bipair neutrino mixing, and also others such as TB, TTB, etc., predict $\sin^2\theta_{13} = 0$, which is inconsistent with the observation.  Theoretically, it is still useful to consider a model giving $\sin^2\theta_{13}=0$, which can be regarded as a reference point to discuss effects of $\sin^2\theta_{13} \neq 0$. It is expected that additional contributions to the mixing angles are produced by the charged lepton contributions if some of the non-diagonal matrix elements of charged lepton mass matrix are non-zero so that the reactor mixing angle can be sifted to the allowed region \cite{theta13}.

In this paper, we explore the ways of minimally modifying the bipair neutrino mixing by including charged lepton contributions. We show that, without arbitrary phases, a minimal modification is realized by just considering one $e$ - $\tau$ contribution from the charged lepton sector in the case 1. The charged lepton contributions give the shift of the reactor mixing angle while the solar and atmospheric mixing angles remain intact. On the other hand, not only $e$ - $\tau$ contribution but also $\mu$ - $\tau$ contribution is required to realize a minimal modification in the case 2. The charged lepton contributions give the shifts of the atmospheric and reactor mixing angles while the solar mixing angle remains intact. 

%%----------------------------------------------------------------------------------
\section{bipair neutrino mixing}
%%----------------------------------------------------------------------------------
We give a brief review of the bipair neutrino mixing scheme. The mixing angles $\theta_{12}$, $\theta_{23}$ and $\theta_{13}$ describe the Pontecorvo-Maki-Nakagawa-Sakata mixing matrix $U$ \cite{UMNS}. In the limit of $\sin^2\theta_{13}$ = $0$, the standard parametrization of the mixing matrix $U$ is given by the Particle Data Group \cite{PDG}, $U=U^0P$ : 
\begin{eqnarray}
U^0&=&\left(
  \begin{array}{ccc}
    c_{12}           & s_{12}          & 0 \\
    -c_{23}s_{12} & c_{23}c_{12}  & s_{23}  \\
    s_{23}s_{12}  & -s_{23}c_{12} & c_{23} 
  \end{array}
\right), 
\end{eqnarray}
where $c_{ij}=\cos\theta_{ij}$ and $s_{ij}=\sin\theta_{ij}$ $(i,j=1,2,3)$. We will also use the notation of $t_{ij}=\tan\theta_{ij}$. The phase matrix $P$ contains two Majorana CP violating phases.

The bipair neutrino mixing $U_{BP}$ is determined by a mixing matrix $U^0$ with two pairs of identical magnitudes of matrix elements. There are two possibilities of the bipair texture \cite{bipair}.

%%----------------------------------------------------------------------------------
\subsection{ Case 1}
%%----------------------------------------------------------------------------------
The first possibility shows $\vert U^0_{12} \vert = \vert U^0_{32} \vert$ and $\vert U^0_{22} \vert = \vert U^0_{23} \vert$. These relations in turn provide useful relationship among the atmospheric neutrino mixing and the solar neutrino mixing as $s_{23}^2 = t_{12}^2$ and $t_{23}^2 = c_{12}^2$. The case 1 of the bipair neutrino mixing is parameterized by only one mixing angle $\theta_{12}$: 
\begin{eqnarray}
U_\nu^{BP1}=\left(
  \begin{array}{ccc}
    c      & s   & 0 \\
    -t^2  & t   & t \\
    st     & -s  & t/c 
  \end{array}
\right), 
\label{Eq:UnuBP1}
\end{eqnarray}
where $c=c_{12}$, $s=s_{12}$ and $t=t_{12}$. The mixing angles are predicted to be $s_{12}^2 = 1 - 1/\sqrt{2} = 0.293$ and $s_{23}^2 = t_{12}^2 = \sqrt{2} - 1 = 0.414$. The case 1 of the bipair neutrino mixing well describes the observed solar and atmospheric neutrino mixing angles.

%%----------------------------------------------------------------------------------
\subsection{ Case 2}
%%----------------------------------------------------------------------------------
The second possibility shows $\vert U^0_{12} \vert = \vert U^0_{22} \vert$ and $\vert U^0_{32} \vert = \vert U^0_{33} \vert$. The atmospheric neutrino mixing is related to the solar neutrino mixing as $c_{23}^2 = t_{12}^2$ and $t_{23}^2 = 1/c_{12}^2$. The case 2 of the bipair neutrino mixing is parameterized by 
\begin{eqnarray}
U_\nu^{BP2}=\left(
  \begin{array}{ccc}
     c      & s   & 0 \\
     -st   & s   & t/c  \\
      t^2  & -t  & t 
  \end{array}
\right).
\label{Eq:UnuBP2}
\end{eqnarray}
The mixing angles are predicted to be $s_{12}^2 = 1 -1/\sqrt{2} = 0.293$ and $s_{23}^2 = 1-t_{12}^2 = 2 - \sqrt{2} = 0.586$. In this case,  although it seems that the atmospheric mixing angle is marginally allowed by the global analysis, the solar and atmospheric mixing angles are consistent with the $2 \sigma$ data. 

%%----------------------------------------------------------------------------------
\section{Charged lepton contributions}
%%----------------------------------------------------------------------------------
The additional contributions to the mixing angles are produced by the charged lepton contributions. The $3 \times 3$ lepton mixing matrix is given by  $U_\ell^\dagger U_\nu$, where $U_\ell$ and $U_\nu$ arise from the diagonalization of the charged lepton mass matrix $M_\ell$ and the neutrino mass matrix $M_\nu$, respectively. These lepton mass matrices satisfy the relations of $M_\ell = U_\ell M_\ell^{diag} U_R^\dagger$ and $M_\nu = U_\nu M_\nu^{diag} U_\nu^T$ where $M_\ell^{diag}$ and $M_\nu^{diag}$ are the diagonal mass matrix of the charged lepton and of the neutrino, $U_R$ transforms between the Lagrangian and the mass basis for the right-handed charged lepton field.

We employ the Dev, Gupta and Gautam (DGG) parametrization \cite{DGG} to reduce the number of complex phases in the mixing matrix to simplify the analysis. A general $3 \times 3$ matrix $U=e^{i\Phi}P\tilde{U}Q$ contains three moduli and six phases, where $P = {\rm diag.}(1, e^{\phi_1}, e^{\phi_2})$ and $Q = {\rm diag.}(1, e^{\rho_1}, e^{\rho_2})$ are phase matrices, $\tilde{U}$ is a matrix containing 3 angles and one phase \cite{generalU}.  In the DGG framework, we consider the following Hermitian product for the charged leptons
\begin{eqnarray}
M_\ell M_\ell^\dagger = U_\ell M_\ell^{diag} U_R^\dagger U_R M_\ell^{diag}U_\ell^\dagger 
=e^{i\phi_\ell}P_\ell \tilde{U}_\ell Q_\ell (M_\ell^{diag})^2Q_\ell^\dagger \tilde{U}_\ell^\dagger P_\ell^\dagger e^{-i\phi_\ell}
=P_\ell \tilde{U}_\ell (M_\ell^{diag})^2 \tilde{U}_\ell^\dagger P_\ell^\dagger.
\end{eqnarray}
Similarly, for the neutrinos, we obtain $M_\nu M_\nu^\dagger = P_\nu \tilde{U}_\nu (M_\nu^{diag})^2 \tilde{U}_\nu^\dagger P_\nu^\dagger$. We can use $P_\ell \tilde{U}_\ell$ for charged leptons and $P_\nu\tilde{U}_\nu$ for neutrinos. As a result, we obtain $U=\tilde{U}_\ell P_\ell^\dagger P_\nu \tilde{U}_\nu$, which contains 6 phases. Three phases from charged leptons can eliminate three of 6 phases, which can be identified with two phases from $P_\ell$ and one phase from $P_\nu$. Since $P_\ell$ is reduced to the unit matrix, the lepton mixing matrix is finally given by
\begin{eqnarray}
\tilde{U} = \tilde{U}_\ell^\dagger \tilde{P} \tilde{U}_\nu,
\end{eqnarray}
where $\tilde{U}_\ell$ and $\tilde{U}_\nu$ contain three real parameters and one phase each while $\tilde{P} = {\rm diag}.(1,1,e^{i\phi})$ contains one phase $\phi$. The mixing matrix $\tilde{U}$ is expressed in terms of six real parameters and three phases.  

The charged lepton mixing matrix $\tilde{U}_\ell$ is parametrized by some manners \cite{theta13}. We adapt the Cabibbo-Kobayashi-Maskawa like parametrization of the matrix $\tilde{U}_\ell$, which is represented by small parameters having magnitude of the order of Wolfenstein parameter $\lambda \sim 0.227$ or less \cite {CKMlikeUell}. In the small angle approximation, the charged lepton mixing matrix becomes
\begin{eqnarray}
\tilde{U}_\ell=\left(
  \begin{array}{ccc}
     1            & \epsilon_{12}   & e^{-i\delta}\epsilon_{13} \\
     -\epsilon_{12}  & 1  &\epsilon_{23}  \\
      -e^{i\delta}\epsilon_{13}       & -\epsilon_{23}  & 1 
  \end{array}	
\right),
\end{eqnarray}
where $\epsilon_{12,13, 23} < 0.227$ and $\delta$ denotes an arbitrary phase.

The neutrino mixing matrix is taken to be $\tilde{U}_\nu = U_\nu^{BP1}$ or $\tilde{U}_\nu = U_\nu^{BP2}$ to study the charged lepton contributions to the mixing angles in the bipair neutrino mixing. As the definition of the bipair neutrino mixing, there is no phase parameter in $U_\nu^{BP1}$ in Eq.(\ref{Eq:UnuBP1}) and $U_\nu^{BP2}$ in Eq.(\ref{Eq:UnuBP2}). The lepton mixing matrix $\tilde{U}$ = $\tilde{U}_\ell \tilde{P} \tilde{U}_\nu^{BP1}$ (and also $\tilde{U}$ = $\tilde{U}_\ell \tilde{P} \tilde{U}_\nu^{BP2}$) has three real parameters $\epsilon_{12,23,13}$ and two phases $\delta, \phi$. The number of parameters in the bipair neutrino mixing is reduced via the conditions of the lepton mixing angles, which are given by 
\begin{eqnarray}
\sin^2\theta_{12} = \frac{\vert \tilde{U}_{12} \vert^2}{\vert \tilde{U}_{11} \vert^2 + \vert \tilde{U}_{12} \vert^2},  \quad
\sin^2\theta_{23} = \frac{\vert \tilde{U}_{23} \vert^2}{\vert \tilde{U}_{23} \vert^2 + \vert \tilde{U}_{33} \vert^2}, \quad 
\sin^2\theta_{13} = \vert \tilde{U}_{13} \vert^2.
\label{Eq:leptonMixingAngles}
\end{eqnarray}
%

%%----------------------------------------------------------------------------------
\subsection{ Case 1 }
%%----------------------------------------------------------------------------------
We take $\tilde{U}_\nu$ = $U_\nu^{BP1}$ and obtain the lepton mixing matrix as 
\begin{eqnarray}
\tilde{U} &=&\left(
  \begin{array}{ccc}
     c    & s  & 0 \\
     -t^2 & t & t  \\
      e^{i\phi} st  & -e^{i\phi} s & e^{i\phi} t/s
  \end{array}
\right) 
+
\left(
  \begin{array}{ccc}
     t^2\epsilon_{12} - \tilde{s} t \epsilon_{13}    & -t \epsilon_{12} + \tilde{s} \epsilon_{13}  & -t \epsilon_{12} -  \frac{\tilde{s}}{c^2}\epsilon_{13} \\
     c\epsilon_{12} -e^{i\phi} st \epsilon_{23} & s\epsilon_{12} + e^{i\phi} s\epsilon_{23} & -e^{i\phi} \frac{t}{c} \epsilon_{23} \\
     e^{i\delta} c\epsilon_{13} - t^2 \epsilon_{23} & e^{i\delta} s\epsilon_{13} + t \epsilon_{23} & t \epsilon_{23}
  \end{array}
\right), \nonumber \\
\end{eqnarray}
where we define $\tilde{s} = e^{-i(\delta - \phi)}s$. From Eq.(\ref{Eq:leptonMixingAngles}), the lepton mixing angles are obtained as 
\begin{eqnarray}
\sin^2\theta_{12} &=& \left( 1-2^{5/4} \epsilon_{12} + 2\epsilon_{13} \cos(\delta - \phi) \right) \sin^2\theta_{12}^{BP1}, \nonumber \\
\sin^2\theta_{23} &=& \left( 1-2^{5/4} \epsilon_{23} \cos\phi \right) \sin^2\theta_{23}^{BP1}, \nonumber \\
\sin^2\theta_{13} &=& \left( \epsilon_{12}^2 + 2^{5/4}\epsilon_{12}\epsilon_{13} \cos(\delta - \phi) \right) \sin^2\theta_{23}^{BP1} 
+ 2 \epsilon_{13}^2 \sin^2\theta_{12}^{BP1}. 
\label{Eq:leptonMixingAnglesInBP1}
\end{eqnarray}
The expressions of $\sin^2\theta_{12}$ and $\sin^2\theta_{23}$ are derived up to the first order of the small $\epsilon_{ij}$. The small parameters $\epsilon_{12}$, $\epsilon_{23}$ and $\epsilon_{13}$ satisfy the following relations
\begin{eqnarray}
\epsilon_{12} &=& \frac{1}{2^{1/4}}\epsilon_{13}\cos(\delta-\phi) + \frac{1}{2^{5/4}} \left( 1 - \frac{1}{r_{12}^{BP1} }\right), \nonumber \\
\epsilon_{23}\cos\phi &=& \frac{1}{2^{5/4}} \left( 1-\frac{1}{r_{23}^{BP1}} \right).
\label{Eq:epsilonsInBP1}
\end{eqnarray}

Since the predicted solar and atmospheric mixing angles are close to the best-fit values of the observed data, to have minimal modification, we assume that the solar and atmospheric mixing angles are not modified by the charged lepton contributions. We take 
\begin{eqnarray}
\sin^2\theta_{12} = \sin^2\theta_{12}^{BP1}, \quad \sin^2\theta_{23} = \sin^2\theta_{23}^{BP1},
\end{eqnarray}
in Eq.(\ref{Eq:leptonMixingAnglesInBP1}) or, equivalently, 
\begin{eqnarray}
\frac{1}{r_{12}^{BP1}} = 1, \quad \frac{1}{r_{23}^{BP1}} = 1,
\end{eqnarray}
in Eq.(\ref{Eq:epsilonsInBP1}) and obtain the following relations 
\begin{eqnarray}
\epsilon_{12} = \frac{1}{2^{1/4}}\epsilon_{13}\cos(\delta-\phi), \quad \epsilon_{23}=0.
\end{eqnarray}
The charged lepton contributions to the mixing angles are parameterized by one real parameter $\epsilon_{13}$ and two phases $\delta, \phi$. 

The mixing angles in the case 1 of the bipair neutrino mixing are obtained as
\begin{eqnarray}
\sin^2\theta_{12} &=& \sin^2\theta_{12}^{BP1}, \nonumber \\
\sin^2\theta_{23} &=& \sin^2\theta_{23}^{BP1}, \nonumber \\
\sin^2\theta_{13} &=& \left(2+\frac{1}{\sqrt{2}}\right)\epsilon_{13}^2\cos^2(\delta -\phi)\sin^2\theta_{23}^{BP1} 
+ 2\epsilon_{13}^2\sin^2\theta_{12}^{BP1}.
\end{eqnarray}
Without arbitrary phases $\delta$ and $\phi$, a minimal modification is realized by just considering one $e$ - $\tau$ contribution from the charged lepton sector, which is denoted by $\epsilon_{13}$ in the case 1 of the bipair neutrino mixing. 

For illustration, we take the arbitrary phases $\delta$ and $\phi$ to satisfy $\cos(\delta - \phi)$ = $1$. If the remaining parameter is taken to be $\epsilon_{13}$ = $0.121$, we obtain $\sin^2\theta_{12}$ = $0.293$, $\sin^2\theta_{23}$ = $0.414$ and $\sin^2\theta_{13}$ = $0.025$ to be  compatible with the best-fit values of the observed data. 

We note that if we use the other best-fit value of the atmospheric mixing $\sin^2\theta_{23} = 0.52$ in the Ref.\cite{otherGlobal} in stead of $\sin^2\theta_{23} = 0.42$ in Eq.(\ref{Eq:bestfit}) from the Ref.\cite{global}, the predicted atmospheric mixing angle is not close to the best-fit value, we can not take $\epsilon_{23}$ = $0$ and the atmospheric mixing angle should be modified by the charged lepton contributions. In this case, to take $\epsilon_{23}\cos\phi$ = $-0.108$ leads to $\sin^2\theta_{23}$ = $0.52$ from Eq.(\ref{Eq:leptonMixingAnglesInBP1}). On the other hand, we can also see in Eq.(\ref{Eq:leptonMixingAnglesInBP1}) that the deviation of the atmospheric mixing angle from the best-fit value depends only on the small $\epsilon_{23}\cos\phi$ while the solar and reactor mixing angles are independent of this $\epsilon_{23}$.

%%----------------------------------------------------------------------------------
\subsection{ Case 2 }
%%----------------------------------------------------------------------------------
We take $\tilde{U}_\nu=U_\nu^{BP2}$ and obtain the lepton mixing matrix as
\begin{eqnarray}
\tilde{U} &=&\left(
  \begin{array}{ccc}
     c    & s  & 0 \\
     -st & s & t/c  \\
      e^{i\phi} t^2  & -e^{i\phi} t & e^{i\phi} t
  \end{array}
\right) 
+
\left(
  \begin{array}{ccc}
     st \epsilon_{12} - \tilde{t} t \epsilon_{13}    & -s \epsilon_{12} + \tilde{t} \epsilon_{13}  & -\frac{t}{c} \epsilon_{12} -  \tilde{t}\epsilon_{13} \\
     c\epsilon_{12} - e^{i\phi}t^2 \epsilon_{23} & s\epsilon_{12} + e^{i\phi} t \epsilon_{23} & -e^{i\phi} t \epsilon_{23}  \\
     e^{i\delta} c\epsilon_{13} - st \epsilon_{23} & e^{i\delta} s \epsilon_{13} + s \epsilon_{23} & \frac{t}{c} \epsilon_{23}
  \end{array}
\right), \nonumber \\
\end{eqnarray}
where we define $\tilde{t} = e^{-i(\delta - \phi)}t$. From Eq.(\ref{Eq:leptonMixingAngles}), the lepton mixing angles are obtained as 
\begin{eqnarray}
\sin^2\theta_{12} &=& \left( 1-2 \epsilon_{12} + 2^{5/4}\epsilon_{13} \cos(\delta - \phi) \right) \sin^2\theta_{12}^{BP2}, \nonumber \\
\sin^2\theta_{23} &=& \left( 1-2^{3/4} \epsilon_{23} \cos\phi \right) \sin^2\theta_{23}^{BP2}, \nonumber \\
\sin^2\theta_{13} &=& \sqrt{2}\left( \epsilon_{13}^2 + 2^{5/4}\epsilon_{12}\epsilon_{13} \cos(\delta - \phi) \right) \sin^2\theta_{12}^{BP2} 
+ \epsilon_{12}^2 \sin^2\theta_{23}^{BP2}.
\label{Eq:leptonMixingAnglesInBP2}
\end{eqnarray}
The expressions of $\sin^2\theta_{12}$ and $\sin^2\theta_{23}$ are derived up to the first order of the small $\epsilon_{ij}$. The small parameters $\epsilon_{12}$, $\epsilon_{23}$ and $\epsilon_{13}$ satisfy the following relations
\begin{eqnarray}
\epsilon_{12} &=& 2^{1/4}\epsilon_{13}\cos(\delta-\phi) + \frac{1}{2}\left( 1 - \frac{1}{r_{12}^{BP2}} \right), \nonumber \\
\epsilon_{23}\cos\phi &=& \frac{1}{2^{3/4}} \left( 1-\frac{1}{r_{23}^{BP2}} \right).
\label{Eq:epsilonsInBP2}
\end{eqnarray}

As same as the case 1, since the predicted solar mixing angles is close to the best-fit value of the observed data, we assume that the solar mixing angle is not modified by the charged lepton contributions in order to have minimal modification. We take 
\begin{eqnarray}
\sin^2\theta_{12} = \sin^2\theta_{12}^{BP2},
\end{eqnarray}
in Eq.(\ref{Eq:leptonMixingAnglesInBP2}) or, equivalently, 
\begin{eqnarray}
\frac{1}{r_{12}^{BP2}} = 1,
\end{eqnarray}
in Eq.(\ref{Eq:epsilonsInBP2}) and obtain the following relation 
\begin{eqnarray}
\epsilon_{12} = 2^{1/4}\epsilon_{13}\cos(\delta-\phi).
\end{eqnarray}
On the other hand, since the predicted atmospheric mixing angle is not close to the best-fit value, we can not take $\epsilon_{23}$ = $0$ and the atmospheric mixing angle should be modified by the charged lepton contributions. The charged lepton contributions to the mixing angles are parameterized by two real parameters $\epsilon_{13}$, $\epsilon_{23}$ and two phases $\delta, \phi$. 

The mixing angles in the case 2 of the bipair neutrino mixing are obtained as 
\begin{eqnarray}
\sin^2\theta_{12} &=& \sin^2\theta_{12}^{BP2}, \nonumber \\
\sin^2\theta_{23} &=& \left( 1-2^{3/4} \epsilon_{23} \cos\phi \right) \sin^2\theta_{23}^{BP2}, \nonumber \\
\sin^2\theta_{13} &=& \sqrt{2}\left(1+2^{3/2} \cos^2(\delta -\phi)\right)\epsilon_{13}^2\sin^2\theta_{12}^{BP2} 
+ \sqrt{2}\epsilon_{13}^2\cos^2(\delta -\phi)\sin^2\theta_{23}^{BP2}.
\end{eqnarray}
Without arbitrary phases $\delta$ and $\phi$, a minimal modification is realized by not only $e$ - $\tau$ contribution but also $\mu$ - $\tau$ contribution, which are denoted by $\epsilon_{13}$ and $\epsilon_{23}$ in the case 2 bipair neutrino mixing. 

For illustration, we take the arbitrary phases to satisfy $\cos(\delta -\phi)$ = $1$. We obtain $\sin^2\theta_{12}$ = $0.293$, $\sin^2\theta_{23}$ = $0.42$, $\sin^2\theta_{13}$ = $0.025$ for $\epsilon_{13}$ = $0.102$ and $\epsilon_{23}\cos\phi$ = $0.168$. These results are well compatible with the best-fit values of the observed data. The case 2 of the bipair neutrino mixing scheme, which was slightly unfavorable, can be favorable if the charged lepton contributions are included.

To compare the predicted reactor neutrino mixing angle $\sin^2\theta_{13}$ in the case 1 of the bipair neutrino mixing with that in the case 2, we show $\sin^2\theta_{13}$ in the case 1 (solid curve) and in the case 2 (dashed curve) as a function of the small parameter $\epsilon_{13}$ in Figure 1. The two horizontal dotted lines correspond to the 2$\sigma$ upper and lower bounds of $\sin^2\theta_{13}$ \cite{global}. We observe that the allowed regions of the small parameter $\epsilon_{13}$ are $0.08 \lesssim \epsilon_{13} \lesssim 0.16$ in the case 1 and $0.07 \lesssim \epsilon_{13} \lesssim 0.13$ in the case 2. 

%--------------------------------------------------------------------
\begin{figure}[t]
\begin{center}
  \includegraphics{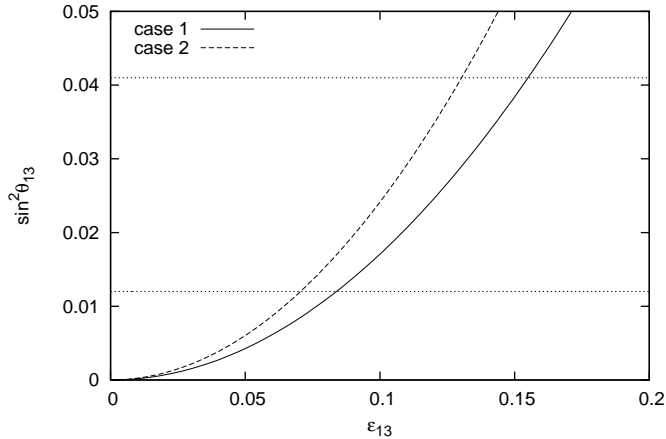}
  \caption{\label{fig:fig1} Predicted reactor neutrino mixing angle $\sin^2\theta_{13}$ as a function of the small parameter $\epsilon_{13}$ in the case 1 (solid curve) and in the case 2 (dashed curve) of the bipair neutrino mixing. The two horizontal dotted lines correspond to the 2$\sigma$ upper and lower bounds of $\sin^2\theta_{13}$ \cite{global}.}
\end{center}
\end{figure}
%--------------------------------------------------------------------

%%----------------------------------------------------------------------------------
\section{Summary}
%%---------------------------------------------------------------------------------- 
%
In the bipair neutrino mixing, the predicted solar and atmospheric neutrino mixing angles are consistent with the observed data while the predicted reactor angle is inconsistent with its observed value. We have explored the ways of minimally modifying the bipair neutrino mixing by including charged lepton contributions. To study the additional contributions to the mixing angles, which are produced by the charged lepton contributions, we have employed the DGG parametrization.

In the case 1 of the neutrino mixing, the predicted solar and atmospheric mixing angles are close to the best-fit values of the observed data. To have minimal modification, we have assumed that the solar and atmospheric mixing angles are not modified by the charged lepton contributions. 
Without arbitrary phases, a minimal modification is realized by just one $e$ - $\tau$ contribution from the charged lepton sector.

In the case 2 of the neutrino mixing, since the predicted solar mixing angles is close to the best-fit value of the observed data, we have assumed that the solar mixing angle is not modified by the charged lepton contributions. On the other hand, since the predicted atmospheric mixing angle is not close to the best-fit value, the atmospheric mixing angle should be modified by the charged lepton contributions. Without arbitrary phases, a minimal modification is realized by not only $e$ - $\tau$ contribution but also $\mu$ - $\tau$ contribution.

%% The Appendices part is started with the command \appendix;
%% appendix sections are then done as normal sections
%% \appendix

%% \section{}
%% \label{}

%% References
%%
%% Following citation commands can be used in the body text:
%% Usage of \cite is as follows:
%%   \cite{key}         ==>>  [#]
%%   \cite[chap. 2]{key} ==>> [#, chap. 2]
%%

%% References with bibTeX database:

\bibliographystyle{elsarticle-num}
\bibliography{<your-bib-database>}

%% Authors are advised to submit their bibtex database files. They are
%% requested to list a bibtex style file in the manuscript if they do
%% not want to use elsarticle-num.bst.

%% References without bibTeX database:

\end{document}